\documentclass[twocolumn]{aastex701}

\usepackage[T1]{fontenc} 
\usepackage[utf8]{inputenc} 
\usepackage{amssymb}
\usepackage{float}
\usepackage{placeins}
\usepackage{hyperref}
\usepackage{natbib}

\setcitestyle{notesep={ }}

\begin{document}

\title{A Cold and Super-Puffy Planet on a Prograde Orbit}

\correspondingauthor{Juan I. Espinoza-Retamal}
\email{jiespinozar@princeton.edu}

\author[0000-0001-9480-8526]{Juan I. Espinoza-Retamal}
\affiliation{Department of Astrophysical Sciences, Princeton University, 4 Ivy Lane, Princeton, NJ 08540, USA}
\affiliation{Instituto de Astrof\'isica, Pontificia Universidad Cat\'olica de Chile, Av. Vicu\~na Mackenna 4860, 7820436 Macul, Santiago, Chile}
\affiliation{Millennium Institute for Astrophysics, Nuncio Monse\~{n}or Sotero Sanz 100, Of. 104, Providencia, Santiago, Chile}
\email{jiespinozar@uc.cl}

\author[0000-0002-9158-7315]{Rafael Brahm}
\affil{Facultad de Ingenier\'ia y Ciencias, Universidad Adolfo Ib\'{a}\~{n}ez, Av. Diagonal las Torres 2640, Pe\~{n}alol\'{e}n, Santiago, Chile}
\affil{Millennium Institute for Astrophysics, Nuncio Monse\~{n}or Sotero Sanz 100, Of. 104, Providencia, Santiago, Chile}
\email{rafael.brahm@uai.cl}

\author[0000-0003-0412-9314]{Cristobal Petrovich}
\affiliation{Department of Astronomy, Indiana University, Bloomington, IN 47405, USA}
\email{cpetrovi@iu.edu}

\author[0000-0002-5389-3944]{Andr\'es Jord\'an}
\affil{Facultad de Ingenier\'ia y Ciencias, Universidad Adolfo Ib\'{a}\~{n}ez, Av. Diagonal las Torres 2640, Pe\~{n}alol\'{e}n, Santiago, Chile}
\affil{Millennium Institute for Astrophysics, Nuncio Monse\~{n}or Sotero Sanz 100, Of. 104, Providencia, Santiago, Chile}
\affiliation{Data Observatory Foundation, Santiago, Chile}
\email{andres.jordan@uai.cl}

\author[0000-0002-1493-300X]{Thomas Henning} %henning@mpia.de (WINE)
\affiliation{Max-Planck-Institut für Astronomie, Königstuhl 17, D-69117 Heidelberg, Germany}
\email{henning@mpia.de}

\author[0000-0002-0236-775X]{Trifon Trifonov} %trifonov@mpia.de 
\affiliation{Max-Planck-Institut für Astronomie, Königstuhl 17, D-69117 Heidelberg, Germany}
\affiliation{Department of Astronomy, Sofia University ``St Kliment Ohridski'', 5 James Bourchier Blvd, BG-1164 Sofia, Bulgaria}
\affiliation{Landessternwarte, Zentrum f\"ur Astronomie der Universit\"at Heidelberg, K\"onigstuhl 12, D-69117 Heidelberg, Germany}
\email{trifonov@mpia.de}

\author[0000-0002-4265-047X]{Joshua N. Winn}
\affiliation{Department of Astrophysical Sciences, Princeton University, 4 Ivy Lane, Princeton, NJ 08540, USA}
\email{jnwinn@princeton.edu}

\author[0009-0009-4849-9764]{Erika Rea}
\affiliation{European Space Agency (ESA), European Space Research and
Technology Centre (ESTEC), Keplerlaan 1, 2201 AZ Noordwijk,
The Netherlands}
\email{rea.erika98@gmail.com}

\author[0000-0002-3164-9086]{Maximilian N. G\"unther} %maximilian.guenther@esa.int
\affiliation{European Space Agency (ESA), European Space Research and
Technology Centre (ESTEC), Keplerlaan 1, 2201 AZ Noordwijk,
The Netherlands}
\email{maximilian.guenther@esa.int}

\author[0000-0001-7948-6493]{Abdelkrim Agabi}
\affiliation{Université Côte d'Azur, Observatoire de la Côte d'Azur, CNRS, Laboratoire Lagrange, CS 34229, F-06304 Nice Cedex 4, France}
\email{karim.agabi@unice.fr}

\author[0000-0002-4278-1437]{Philippe Bendjoya}
\affiliation{Université Côte d'Azur, Observatoire de la Côte d'Azur, CNRS, Laboratoire Lagrange, CS 34229, F-06304 Nice Cedex 4, France}
\email{bendjoya@oca.eu}

\author[0000-0002-5181-0463]{Hareesh Bhaskar} 
\affiliation{Department of Astronomy, Indiana University, Bloomington, IN 47405, USA}
\email{bhareeshg@gmail.com}

\author[0000-0002-7613-393X]{François Bouchy} 
\affiliation{Observatoire de Genève, Département d'Astronomie, Université de Genève, Chemin Pegasi 51b, 1290 Versoix, Switzerland}
\email{Francois.Bouchy@unige.ch}

\author[0000-0001-6003-8877]{M\'arcio Catelan} % mcatelan@uc.cl 
\affiliation{Instituto de Astrof\'isica, Pontificia Universidad Cat\'olica de Chile, Av. Vicu\~na Mackenna 4860, 7820436 Macul, Santiago, Chile}
\affiliation{Millennium Institute for Astrophysics, Nuncio Monse\~{n}or Sotero Sanz 100, Of. 104, Providencia, Santiago, Chile}
\email{mcatelan@uc.cl}

\author[0000-0002-9196-5734]{Carolina Charalambous}
\affiliation{Instituto de Astrof\'isica, Pontificia Universidad Cat\'olica de Chile, Av. Vicu\~na Mackenna 4860, 7820436 Macul, Santiago, Chile}
\email{caritocharalambous@gmail.com}

\author[0009-0007-5876-546X]{Vincent Deloupy}
\affiliation{École Normale Supérieure, Département de Physique, Rue d'Ulm, 75005 Paris Cedex 5, France}
\email{vincent.deloupy@gmail.com}

\author[0000-0002-3937-630X]{George Dransfield}
\affiliation{School of Physics \& Astronomy, University of Birmingham, Edgbaston, Birmingham B15 2TT, UK}
\affiliation{Department of Astrophysics, University of Oxford, Denys Wilkinson Building, Keble Road, Oxford OX1 3RH, UK}
\affiliation{Magdalen College, University of Oxford, Oxford OX1 4AU, UK}
\email{george.dransfield@magd.ox.ac.uk}

\author[0000-0003-3130-2768]{Jan Eberhardt} %eberhardt@mpia.de (WINE)
\affiliation{Max-Planck-Institut für Astronomie, Königstuhl 17, D-69117 Heidelberg, Germany}
\email{eberhardt@mpia.de}

\author[0000-0001-9513-1449]{N\'estor Espinoza} %nespinoza@stsci.edu (WINE)
\affiliation{Space Telescope Science Institute, 3700 San Martin Drive, Baltimore, MD 21218, USA}
\email{nespinoza@stsci.edu}

\author[0009-0007-1053-0004]{Alix V. Freckelton} %axf859@student.bham.ac.uk
\affiliation{School of Physics \& Astronomy, University of Birmingham, Edgbaston, Birmingham B15 2TT, UK}
\email{axf859@student.bham.ac.uk}

\author[0000-0002-7188-8428]{Tristan Guillot}
\affiliation{Université Côte d'Azur, Observatoire de la Côte d'Azur, CNRS, Laboratoire Lagrange, CS 34229, F-06304 Nice Cedex 4, France}
\email{tristan.guillot@oca.eu}

\author[0000-0002-5945-7975]{Melissa J. Hobson} %melihobson@gmail.com (WINE)
\affiliation{Observatoire de Genève, Département d'Astronomie, Université de Genève, Chemin Pegasi 51b, 1290 Versoix, Switzerland}
\email{melihobson@gmail.com}

\author{Mat\'ias I. Jones} %mjones@eso.org (WINE)
\affiliation{European Southern Observatory (ESO), Alonso de C\'ordova 3107, Vitacura, Casilla 19001, Santiago, Chile}
\email{mjones@eso.org}

\author[0000-0001-9699-1459]{Monika Lendl}
\affiliation{Observatoire de Genève, Département d'Astronomie, Université de Genève, Chemin Pegasi 51b, 1290 Versoix, Switzerland}
\email{Monika.Lendl@unige.ch}

\author[0000-0001-5000-7292]{Djamel Mekarnia}
\affiliation{Université Côte d'Azur, Observatoire de la Côte d'Azur, CNRS, Laboratoire Lagrange, CS 34229, F-06304 Nice Cedex 4, France}
\email{djamel.mekarnia@oca.eu}

\author[0000-0003-2186-234X]{Diego J. Muñoz} 
\affiliation{Department of Astronomy and Planetary Science, Northern Arizona University, Flagstaff, AZ 86011, USA}
\email{diego.munoz.anguita@gmail.com}

\author[0000-0002-5254-2499]{Louise D. Nielsen}
\affiliation{Observatoire de Genève, Département d'Astronomie, Université de Genève, Chemin Pegasi 51b, 1290 Versoix, Switzerland}
\affiliation{University Observatory, Faculty of Physics, Ludwig-Maximilians-Universität München, Scheinerstr. 1, 81679 Munich, Germany}
\email{Louise.Nielsen@lmu.de}

\author[0000-0003-3047-6272]{Felipe I. Rojas} %firojas@uc.cl (WINE)
\affiliation{Instituto de Astrof\'isica, Pontificia Universidad Cat\'olica de Chile, Av. Vicu\~na Mackenna 4860, 7820436 Macul, Santiago, Chile}
\email{firojas@uc.cl}

\author[0000-0003-3914-3546]{François-Xavier Schmider}
\affiliation{Université Côte d'Azur, Observatoire de la Côte d'Azur, CNRS, Laboratoire Lagrange, CS 34229, F-06304 Nice Cedex 4, France}
\email{schmider@oca.eu}

\author[0000-0002-7444-5315]{Elyar Sedaghati}
\affiliation{European Southern Observatory (ESO), Alonso de C\'ordova 3107, Vitacura, Casilla 19001, Santiago, Chile}
\email{esedagha@eso.org}

\author[0000-0001-7409-5688]{Guðmundur Stefánsson} 
\affil{Anton Pannekoek Institute for Astronomy, University of Amsterdam, Science Park 904, 1098 XH Amsterdam, The Netherlands}
\email{g.k.stefansson@uva.nl}

\author[0009-0008-5145-0446]{Stephanie Striegel} 
\affil{SETI Institute, Mountain View, CA 94043 USA}
\affil{NASA Ames Research Center, Moffett Field, CA 94035 USA}
\email{stephanie.striegel@nasa.gov}

\author[0000-0002-3503-3617]{Olga Suarez}
\affiliation{Université Côte d'Azur, Observatoire de la Côte d'Azur, CNRS, Laboratoire Lagrange, CS 34229, F-06304 Nice Cedex 4, France}
\email{olga.suarez@oca.eu}

\author[0009-0004-8891-4057]{Marcelo Tala Pinto}
\affil{Department of Astronomy, McPherson Laboratory, The Ohio State University, 140 W 18th Ave, Columbus, Ohio 43210, USA}
\email{tala.1@osu.edu}

\author[0009-0008-2214-5039]{Mathilde Timmermans}
\affiliation{School of Physics \& Astronomy, University of Birmingham, Edgbaston, Birmingham B15 2TT, UK}
%\affiliation{Astrobiology Research Unit, University of Liège, Allée du 6 août, 19, 4000 Liège (Sart-Tilman), Belgium}
\email{m.timmermans@bham.ac.uk}

\author[0000-0002-5510-8751]{Amaury H. M. J. Triaud}
\affiliation{School of Physics \& Astronomy, University of Birmingham, Edgbaston, Birmingham B15 2TT, UK}
\email{a.triaud@bham.ac.uk}

\author[0000-0001-7576-6236]{Stéphane Udry}
\affiliation{Observatoire de Genève, Département d'Astronomie, Université de Genève, Chemin Pegasi 51b, 1290 Versoix, Switzerland}
\email{Stephane.Udry@unige.ch}

\author[0000-0003-2417-7006]{Solène Ulmer-Moll}
\affiliation{Leiden Observatory, Leiden University, P.O. Box 9513, 2300 RA Leiden, The Netherlands}
\affiliation{Observatoire de Genève, Département d'Astronomie, Université de Genève, Chemin Pegasi 51b, 1290 Versoix, Switzerland}
\email{ulmer-moll@strw.leidenuniv.nl}

\author[0000-0002-0619-7639]{Carl Ziegler}
\affiliation{Department of Physics, Engineering and Astronomy, Stephen F. Austin State University, 1936 North St, Nacogdoches, TX 75962, USA}
\email{carl.ziegler@sfasu.edu}

\begin{abstract}
We report the discovery of TOI-4507 b, a transiting sub-Saturn with a density $<$\,0.2~g/cm$^3$ on a 105-day prograde orbit around a 700~Myr old F star. The transits were detected using data from TESS as well as the Antarctic telescope ASTEP. A joint analysis of the light curves and radial velocities from HARPS, FEROS, and CORALIE confirmed the planetary nature of the signal by limiting the mass to be below $20\,M_\oplus$ at 95\% confidence. The radial velocities also exhibit the Rossiter-McLaughlin effect and imply that the planet orbits the star in a prograde orbit with a sky-projected obliquity $\lambda=-15_{-44}^{+50}$ deg ($|\lambda|<80$ deg at $3\sigma$). With these characteristics, TOI-4507 is one of the longest-period systems for which the stellar obliquity has been measured, and the planet is among the longest-period and youngest ``super-puff'' planets yet discovered.
\end{abstract}


\section{Introduction} 

More than 6,000 exoplanets have been discovered\footnote{\url{https://exoplanetarchive.ipac.caltech.edu/}}, most of them using the methods of transits and radial velocities (RVs). The {\it Kepler} \citep{Borucki2010} and Transiting Exoplanet Survey Satellite \citep[TESS;][]{Ricker2015} discoveries have revealed valuable information about the demographics of short-period exoplanets \citep[see, e.g., a review by][]{Zhu2021}. However, because the probability of observing transits decreases with orbital distance, it is challenging to explore the outer regions of planetary systems using transits \citep[e.g.,][]{Gaudi2005}. Likewise, RV surveys have revealed valuable information about the demographics of cold Jupiters \citep[e.g.,][]{Cumming2008,Wittenmyer2016,Fernandes2019,Bonomo23}, but because the RV semiamplitude decreases with decreasing planetary mass and increasing semimajor axis, it has been difficult to study less massive cold worlds.

Interestingly, recent discoveries have uncovered the existence of the so-called ``super-puff'' exoplanets \citep[e.g.,][]{Masuda2014,Santerne2019,Barkaoui2024}, which have sizes comparable to that of Jupiter but masses less than that of Saturn or even comparable to that of Neptune, leading to mean densities $\lesssim0.3$~g/cm$^3$. Possible explanations for these low densities involve the accretion of unusually thick H/He envelopes under conditions that enable rapid cooling, such as dust-free formation beyond 1 au followed by inward migration \citep[e.g.,][]{Lee2016} or inflation due to tidal dissipation \citep[e.g.,][]{Millholland2019,Millholland2020,Sethi2025}. It is also possible that some planets with apparently low densities are actually normal planets with large opaque rings \citep[e.g.,][]{Akinsanmi2020,Saillenfest2023,Lu2025}, a hypothesis that has been explored in detail for HIP-41378~f \citep{Vanderburg2016,Santerne2019}.

In this letter, we present the discovery and characterization of TOI-4507 b, a cold planet transiting a young F star with an orbital period of 105 days. Our observations reveal that the planet has an unusually low density, making it one of the longest-period super-puffs known to date. Furthermore, our analysis of the Rossiter-McLaughlin (RM) effect \citep{Rossiter1924,McLaughlin1924} shows that the planet has a prograde orbit, making it one of the longest-period systems for which the stellar obliquity has been measured. A more detailed validation of TOI-4507 b using TESS and ASTEP data appears in a separate paper \citep{Rea2025}.

\section{Observations}\label{sec:observations}

\subsection{Photometry}

\subsubsection{TESS}

\begin{figure*}[t!]
    \centering
    \includegraphics[width=\textwidth]{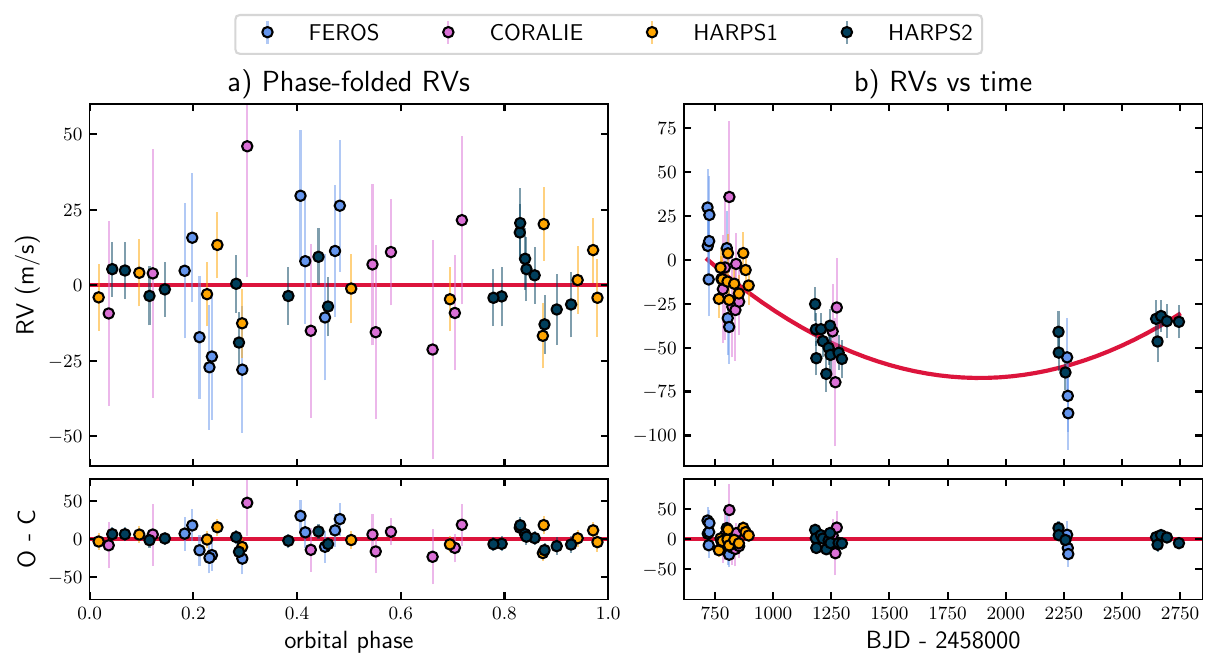}
    \includegraphics[width=\textwidth]{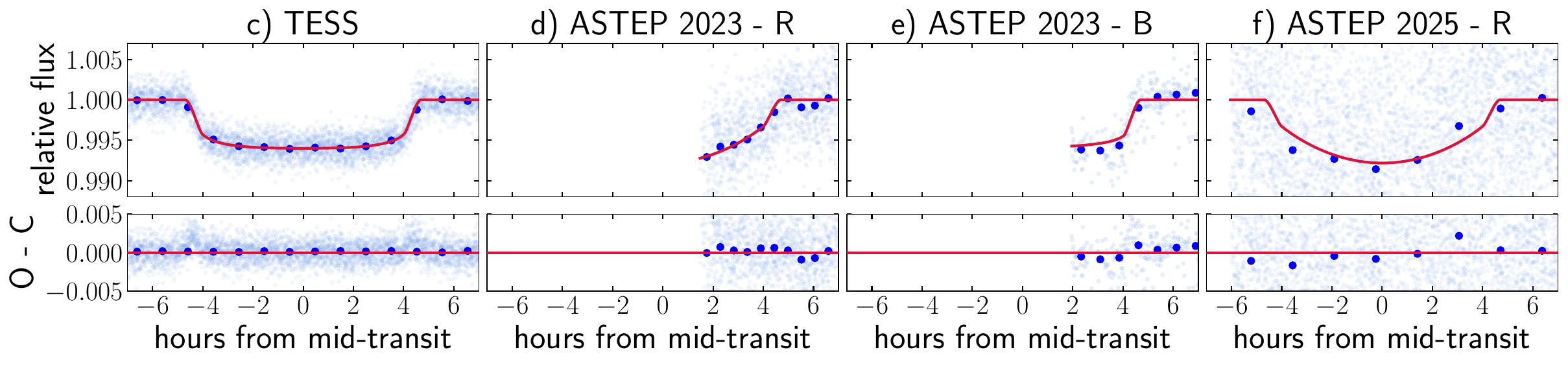}
    \caption{Different observations of TOI-4507 along with the best model. a) Phase-folded out-of-transit RVs taken with different instruments, along with the best model in red. b) RVs as a function of time showing the long-term trend detected at a $5\sigma$ confidence level. c-f) Different light curves along with the best model. Binned data are shown as solid blue points.}
    \label{fig:TOI-4507}
\end{figure*}

Between 2018 and 2024, TOI-4507 was observed by TESS in Sectors 2, 3, $5-13$, $27-30$, $32-39$, $61-69$, and $87-89$, and all of the data are available with 2-minute time sampling. The TESS Science Processing Operations Center \citep[SPOC;][]{spoc} pipeline identified a transit signal with a periodicity of 105 days and a depth of 6,100 ppm. We downloaded and combined all the
available light curves using the code provided by the \citet{lightkurve}. We used the Presearch Data Conditioning Simple Aperture Photometry SPOC light curves \citep{Smith2012,Stumpe2012,Stumpe2014}, which are corrected for pointing and focus-related instrumental signatures, discontinuities resulting from radiation events in the CCD detectors, outliers, and contributions to the recorded flux from nearby stars. The TESS light curve, along with the best transit model, is shown in Figure~\ref{fig:TOI-4507}.

\subsubsection{ASTEP}

We followed up on TESS observations using photometry from the ground. We observed one transit egress on May 23, 2023, and one full transit on May 24, 2025, using the Antarctic Search for Transiting ExoPlanets (ASTEP) 0.4 m telescope \citep{Guillot2015,Mekarnia2016} located at Concordia station in Antarctica. Observations were performed simultaneously in the $R$ and $B$ bands, however, the observations of the 2025 transit are only available in the $R$ band due to a technical issue. ASTEP data were processed on-site using a combination of \texttt{IDL} and \texttt{python} aperture photometry pipelines \citep[e.g.,][]{Abe2013,Mekarnia2016,Dransfield2022}. The resulting light curves, along with the best models, are shown in Figure~\ref{fig:TOI-4507}. A more detailed validation of TOI-4507 b using TESS and ASTEP data is presented in a separate paper \citep{Rea2025}.

\subsection{Spectroscopy}

\subsubsection{HARPS}

In order to confirm the planetary nature of TOI-4507 b, we observed the target with the High Accuracy Radial velocity Planet Searcher (HARPS) spectrograph \citep{Mayor2003} installed on the ESO 3.6 m telescope at La Silla observatory, Chile. HARPS is an echelle spectrograph with a resolving power of $R\approx115,\!000$, which covers the wavelength range of $380-690$ nm. We obtained a total of 31 out-of-transit spectra between October 2019 and March 2025 in the context of the Warm gIaNts with tEss (WINE) collaboration \citep[e.g.,][]{Brahm2023,Tala2025,Eberhardt2025,Vitkova2025}, whose main objective is to detect and characterize relatively long-period transiting exoplanets. The exposure times range between 1200 s and 1500 s, contingent upon prevailing sky and seeing conditions.

In addition to the out-of-transit HARPS observations, we observed about 70\% of a transit of TOI-4507 b on October 27, 2024, between 02:18 and 08:48 UTC. We obtained 20 spectra of the host star during the transit, with an exposure time of 1200 s. The observations were performed under clear sky conditions, with a median atmospheric seeing of $1.3^{\prime\prime}$ and airmass varying from 2.0 to 1.3.

The spectra were reduced using the \texttt{ceres} pipeline \citep{ceres}, which, starting from the raw images, performs all reduction steps and derives precise RVs via the cross-correlation function (CCF) technique. The median signal-to-noise ratio (S/N) of the processed spectra is 50 per resolution element at 550 nm, leading to a median formal RV uncertainty of 4.7 m/s. Because of the warm-up of HARPS on March 23, 2020, due to the COVID-19 pandemic, the data taken before and after this date were considered as having come from different instruments (HARPS1 and HARPS2) with different additive RV offsets and ``jitter'' parameters (the level of uncertainty in excess of the formal uncertainty). The HARPS RVs of the target, together with the best model, are shown in Figure~\ref{fig:TOI-4507} and available in Appendix~\ref{app:rvs}.

\subsubsection{FEROS}

We also observed the target using the Fiber-fed Extended Range Optical Spectrograph \cite[FEROS;][]{Kaufer1999} installed in the MPG/ESO 2.2 m telescope in La Silla Observatory in Chile. FEROS is an echelle spectrograph with a resolving power of $R\approx48,\!000$ and a wavelength range coverage of $350-920$ nm. We obtained a total of 11 out-of-transit spectra of the star between August 2019 and November 2023. The exposure times range between 600 s and 1800 s. FEROS observations were also processed using the \texttt{ceres} pipeline and RVs extracted using the CCF method. The median S/N of the FEROS spectra is 95 per resolution element at 550 nm, with a median RV uncertainty of 9.5 m/s. The resulting FEROS RVs for TOI-4507, along with the best-fit model, are presented in Figure~\ref{fig:TOI-4507} and available in Appendix~\ref{app:rvs}.

\subsubsection{CORALIE}

We also observed the target using the CORALIE spectrograph \citep{Queloz2001} installed in the Swiss 1.2~m Leonhard Euler Telescope at ESO’s La Silla Observatory. CORALIE is an echelle spectrograph with a resolving power of $R\approx60,\!000$, which covers the wavelength range of $390-680$ nm. We observed TOI-4507 using exposure times varying between 900 and 1800 s, together with the simultaneous Fabry-Pérot étalon. We obtained a total of 10 out-of-transit spectra between October 2019 and February 2021. Data reduction was done using the dedicated CORALIE data reduction pipeline, which also derives RV using the CCF technique and a G2 mask. CORALIE observations have a median RV uncertainty of 28 m/s and are shown in Figure~\ref{fig:TOI-4507} along with the best model. Data is available in Appendix~\ref{app:rvs}.

\subsection{High-Resolution Imaging}

\begin{figure}
    \centering
    \includegraphics[width=\columnwidth]{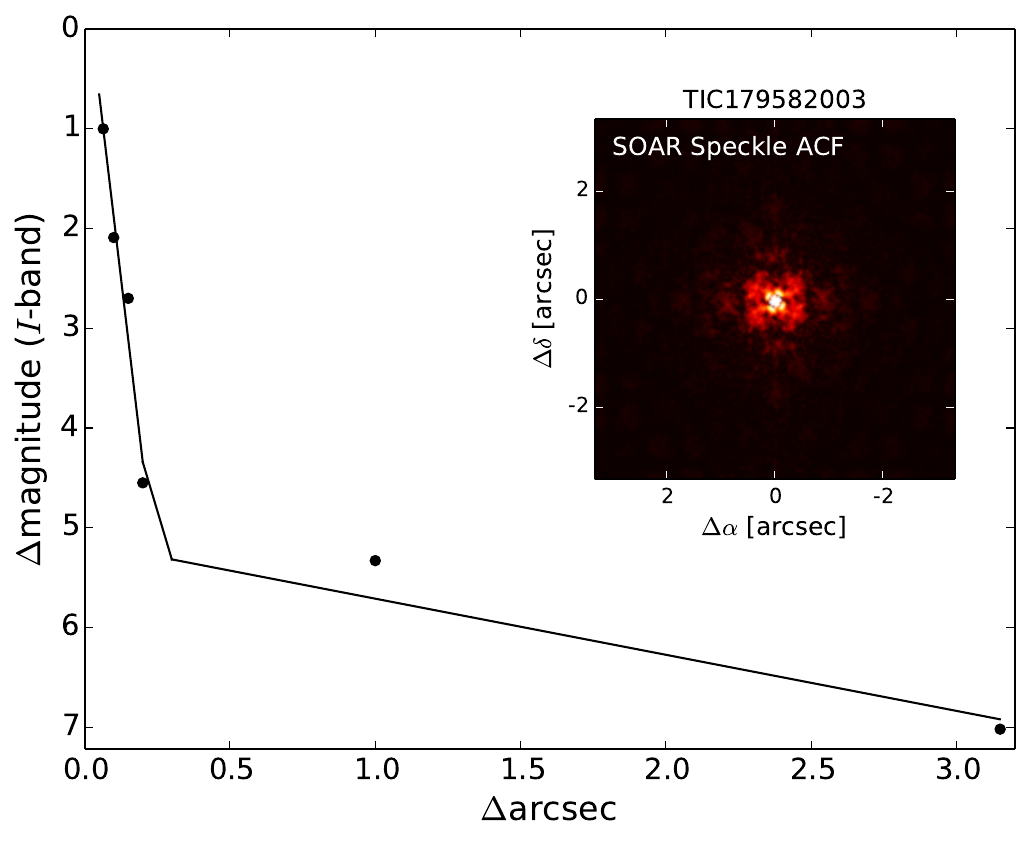}
    \caption{High-resolution imaging observations of TOI-4507 taken with the 4.1 m SOAR telescope. We show the $5\sigma$ detection sensitivity and speckle autocorrelation functions. No companions to TOI-4507 within $3^{\prime\prime}$ are found in these observations.}
    \label{fig:soar}
\end{figure}

In order to search for nearby sources that may contaminate the TESS photometry, resulting in an underestimated planetary radius, or act as astrophysical false positives, such as background eclipsing binaries, we took a high-resolution image of the target. We observed TOI-4507 with the SOuthern Astrophysical Research (SOAR) 4.1 m telescope \citep{Sebring2003,Tokovinin2018} on the night of November 20, 2021. This observation was performed using an $I$ filter and is sensitive enough to detect a 7 mag fainter star at an angular distance of $1^{\prime\prime}$ from the target. More details of the observations within the SOAR--TESS survey are available in \citet{Ziegler2020}. The $5\sigma$ detection sensitivity and speckle autocorrelation functions from the observations are shown in Figure~\ref{fig:soar}. No nearby stars were detected within $3^{\prime\prime}$ of TOI-4507 in the SOAR observations.

\section{Stellar characterization}\label{sec:stellar}

To estimate the stellar parameters of TOI-4507, we followed the two-step iterative procedure presented in \citet{Brahm2019}. Briefly, in the first step, we computed the stellar atmospheric parameters using the \texttt{zaspe} package \citep{zaspe}, which compares the co-added high-resolution HARPS spectrum with a grid of synthetic ones to determine the best fit. The search is performed in the spectral regions that are most sensitive to changes in the stellar parameters, and reliable error bars are computed through Monte Carlo simulations. In the second step, we computed the stellar physical parameters by fitting stellar evolutionary models to the observed spectral energy distribution. We fitted public broadband photometric data to synthetic magnitudes generated from the \texttt{PARSEC} isochrones \citep{parsec} and adopting the \textit{Gaia} Data Release (DR) 3 parallax \citep{Gaia_DR3}. We model interstellar extinction using the prescription of \citet{cardelli}. In this step, the stellar temperature derived with \texttt{zaspe} is used as a prior, while the metallicity is held fixed. From the stellar mass and radius obtained with the second step, we obtained a more precise value of $\log{g}$, which is held fixed in a new iteration of the first step. This procedure is repeated until reaching convergence in $\log{g}$. The resulting parameters of TOI-4507 are presented in Table \ref{tab:stellar}.

\begin{deluxetable*}{llcr}
\tablecaption{Stellar properties$^a$ of TOI-4507.  \label{tab:stellar}}
\tablecolumns{4}
\tablewidth{0pt}
\tablehead{Parameter & Description & Value & Reference}
\startdata
RA & Right Ascension (J2015.5) & 05h21m48.33s & {\it Gaia} DR3$^b$\\
Dec & Declination (J2015.5) & -69d59m17.58s & {\it Gaia} DR3\\
pm$^{\rm RA}$ & Proper motion in RA (mas/yr) & 22.00$\pm$0.02 & {\it Gaia} DR3\\
pm$^{\rm Dec}$ & Proper motion in DEC (mas/yr) & 3.97$\pm$0.02 & {\it Gaia} DR3\\
$\pi$ & Parallax (mas) & 5.64$\pm$0.01 &  {\it Gaia} DR3 \\
$d$ & Distance (pc) & 177.3$\pm$0.3 & {\it Gaia} DR3 \\
\hline
T & TESS magnitude (mag) & 11.230$\pm$0.007 & TICv8$^c$\\
B  & B-band magnitude (mag) & 11.52$\pm$0.04 & APASS$^d$\\
V  & V-band magnitude (mag) &  10.806$\pm$0.030 & APASS\\
G  & {\it Gaia} G-band magnitude (mag) & 10.567$\pm$0.002 & {\it Gaia} DR3\\
G$_{\rm BP}$ & {\it Gaia} BP-band magnitude (mag) &  10.848 $\pm$ 0.005  & {\it Gaia} DR3\\
G$_{\rm RP}$ & {\it Gaia} RP-band magnitude (mag) &  10.133 $\pm$ 0.003  & {\it Gaia} DR3\\
J & 2MASS J-band magnitude (mag) & 9.69 $\pm$ 0.02  & 2MASS$^e$\\
H & 2MASS H-band magnitude (mag) & 9.43 $\pm$ 0.02  &  2MASS\\
K$_s$ & 2MASS K$_s$-band magnitude (mag) & 9.38 $\pm$0.03 & 2MASS\\
\hline
$T_{\rm eff}$ & Effective temperature (K) & 6260 $\pm$ 80 & This work\\
$\log{g}$ & Surface gravity (cgs) & 4.45 $\pm$ 0.02    & This work\\
$[$Fe/H$]$ & Metallicity (dex) & -0.06 $\pm$ 0.05  & This work\\
$v\sin{i_\star}$ & Projected rotational velocity (km/s) & 4.6 $\pm$ 0.9   & This work\\
$M_{\star}$ & Mass ($M_\odot$) & 1.11 $\pm$ 0.02 & This work\\
$R_{\star}$ & Radius ($R_\odot$) & 1.04 $\pm$ 0.01  & This work\\
L$_{\star}$ & Luminosity ($L_\odot$) & 1.41 $\pm$ 0.05 & This work\\
A$_{V}$ & Visual extinction (mag) & 0.07 $\pm$ 0.04 & This work\\
Age & Age (Gyr) & 0.7$_{-0.5}^{+0.8} $ & This work\\
$\rho_\star$ & Density (g/cm$^{3}$) & 1.40$_{-0.05}^{+0.04}$ & This work\\
\enddata
\tablecomments{$^a$ The stellar parameters computed in this work do not consider possible systematic differences among different stellar evolutionary models \citep{tayar:2022} and have underestimated uncertainties.\\
$^b$\citet{Gaia_DR3}.\\
$^c$ \citet{Stassun2018,Stassun2019}. The TESS magnitude is shown only for reference and was not included in our stellar analysis.\\
$^d$ \citet{apass}.\\
$^e$ \citet{2mass}.}
\end{deluxetable*}

\section{Photometric analysis}\label{sec:phot}

To determine the transit ephemerides of TOI-4507 b and look for transit timing variations (TTVs), we analyzed the photometry presented in Section \ref{sec:observations} with the \texttt{juliet} code \citep{juliet}. This code uses \texttt{batman} \citep{batman} for the transit model and the \texttt{dynesty} dynamic nested sampler \citep{dynesty} to sample the posteriors. We placed uniform priors on the impact parameter $b$ and radius ratio $R_p/R_{\star}$, with a Gaussian prior on the stellar density $\rho_\star$ that was constrained in Section \ref{sec:stellar}. We sampled the limb darkening parameters using the quadratic $q_1$ and $q_2$ parameters from \citet{Kipping13} with uniform priors. We placed Gaussian priors for each transit mid-point based on the expected values calculated from the orbital period and time of mid-transit from the SPOC solution, placing a large width of 1 day on the prior to not impact the derived transit midpoints. To account for variability and systematic noise in the TESS light curve, we included a Matern-3/2 Gaussian Process (GP) as implemented in \texttt{celerite} \citep{celerite} and available in \texttt{juliet}. From this analysis, we found $P = 104.61573\pm0.00008$ d and $t_0 = 2458413.8572\pm0.0007$ (BJD). The nearly continuous TESS time series between Sectors 5 and 13, 32 and 39, and 61 and 69 left no doubt that the true orbital period is about 105 days, as consecutive transits are seen in the TESS data. Here, we also obtained a detrended TESS light curve that was used in the subsequent analyses. Finally, we also found tentative evidence for TTVs, as the transit observed with ASTEP in 2025 deviates by $\sim25$ minutes from the linear ephemerides solution. The rest of the transits do not show this behavior, being in all cases within $\sim5$ minutes of the linear solution.

The TESS light curve exhibits a quasi-periodic modulation with a periodicity of $\sim1.7$ days. If this were due to stellar rotation, then the combination of the measurements of rotation period, stellar radius, and projected rotation velocity would imply a low value of $\sin i_\star \approx 0$ and a nearly-polar orbit for the planet. This was our initial interpretation. However, a more detailed analysis revealed that the 1.7-day variability comes from a blended variable star that lies in the background (see Appendix \ref{app:contamination}). Without an estimate of the rotation period of TOI-4507, we cannot constrain the inclination of the stellar rotation axis.

\section{Joint Photometric and RV Modeling}

In order to derive the parameters of the planet, we performed a joint analysis of all the available photometry and out-of-transit RVs using the \texttt{ironman}\footnote{\url{https://github.com/jiespinozar/ironman}} code \citep{Espinoza-Retamal2023b,Espinoza-Retamal2024}. In brief, \texttt{ironman} is a \texttt{python} code that can jointly fit in-transit and out-of-transit RVs together with transit photometry. To model the photometry and RVs, \texttt{ironman} uses \texttt{radvel} \citep{radvel} and \texttt{batman} \citep{batman}, respectively. Finally, to get the posteriors \texttt{ironman} uses the \texttt{dynesty} dynamic nested sampler \citep{dynesty}. A copy of \texttt{ironman} version 1.1.0, which was used in the analysis, is available in Zenodo \citep[see][]{zenodo}.

To reduce the computational cost, we only considered the detrended TESS data within 7.5 hours of a transit midpoint. We included independent jitter terms for each RV and photometric instrument to account for possible instrumental systematics. We placed uniform priors on all of the parameters except for the stellar density, for which we used the constraint described in Section \ref{sec:stellar}. Finally, to account for the possible non-detection of the Keplerian signal in the RVs, we allowed the semiamplitude to take negative values. All priors and resulting posteriors are shown in Table~\ref{tab:fit}.

Figure \ref{fig:TOI-4507} shows the different datasets along with the best model. We found that TOI-4507 b is comparable to Saturn in size, having a radius of $8.2\pm0.1\,R_\oplus$. However, the Keplerian signal was not detected. We found $K$ consistent with 0 m/s, leading to a 2$\sigma$ upper limit on the planet's mass of $20\,M_\oplus$, comparable to that of Neptune. This combination results in a bulk density $<0.2$ g/cm$^3$, making TOI-4507 b a super-puff planet and one of the longest-period planets known to be in this category. Future RV monitoring with more precise instruments will be necessary to precisely measure the mass and density of the planet.

There is a hint of a small eccentricity, $e=0.09^{+0.26}_{-0.06}$. Indeed, we fitted the data using both a circular and an eccentric model. For the eccentric case, we sampled the eccentricity and argument of periastron as $\sqrt{e}\cos{\omega}$ and $\sqrt{e}\sin{\omega}$. The Bayesian evidence difference ($\Delta\log{Z}$), being $<2$, is not enough to favor any of the models in particular. As both models returned a fully consistent set of parameters, we elected to formally adopt the values from the eccentric fit to highlight the possible range of eccentricities compatible with the data. This possible eccentricity can also be further constrained with additional RV measurements.

We found evidence for a long-term RV trend, possibly caused by a longer-period outer companion. We modeled the trend using a quadratic model, $\ddot{\gamma}(t-t_a)^2+\dot{\gamma}(t-t_a)$, where $t_a$ was arbitrarily chosen to be the time of the earliest precise RV measurement. The fit gave $\ddot{\gamma}=0.0180\pm0.0035$ m/s/day/yr and $\dot{\gamma}=-0.12\pm0.02$ m/s/day, implying a $5\sigma$ detection. Further information about this possible companion, including its orbital inclination, might be possible by combining additional RVs and {\it Gaia} DR4 astrometry \citep[e.g.,][]{Espinoza-Retamal2023a}, which would help to understand the system's dynamical history.

\section{Rossiter-McLaughlin Effect}\label{sec:RM}

\begin{figure}[t!]
    \centering
    \includegraphics[width=\columnwidth]{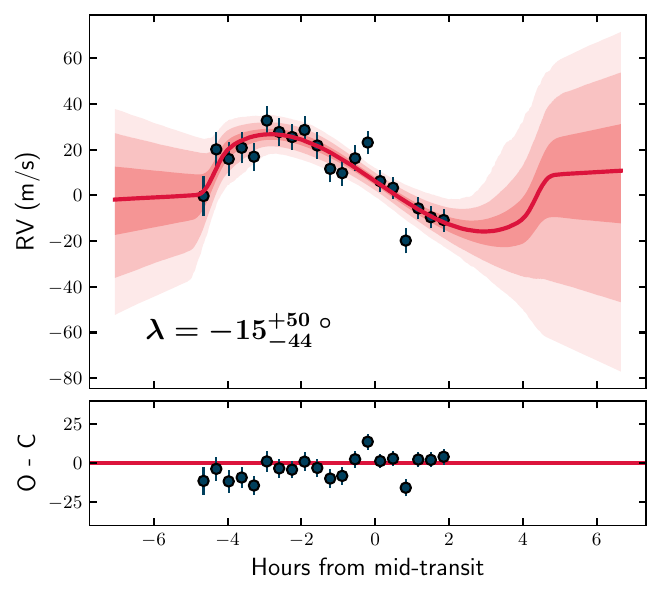}
    \caption{HARPS observations of the RM effect produced during the transit of TOI-4507 b, along with the best-fit model in red. Shaded areas show $1\sigma$, $2\sigma$, and $3\sigma$ models.}
    \label{fig:RM}
\end{figure}

In order to constrain the stellar obliquity, we analyzed the HARPS observations of the RM effect using \texttt{ironman}. To model the RM effect, \texttt{ironman} uses \texttt{rmfit} \citep{Stefansson20,Stefansson22}, which implements the equations from \citet{Hirano10}. Given the lack of an estimation of the rotational period of TOI-4507 (see Appendix \ref{app:contamination}), here we only derived the sky-projected obliquity by directly sampling $\lambda$ and $v\sin{i_\star}$, both with uninformative uniform priors. We placed informative priors for almost all the rest of the parameters based on the values reported in Table \ref{tab:fit}. We also placed an informative prior on $\beta$, the intrinsic linewidth accounting for instrumental and macroturbulence broadening. We considered an instrumental broadening of 2.6~km/s because of the HARPS resolution, and for the macroturbulence broadening, we followed \citet{Albrecht12} and used the macroturbulence law for hot stars from \citet{Gray84} considering $T_{\rm eff}=6260$ K, which resulted in 5.5~km/s. We added the instrumental and macroturbulence broadening in quadrature to set our prior, with an uncertainty of 2 km/s. Finally, to account for possible RV trends in the RM data due to orbital motion or stellar activity, we allowed for an arbitrary RV slope in the model. All priors and resulting posteriors are shown in Table \ref{tab:RM}.

As shown in Figure \ref{fig:RM}, we found that TOI-4507 b has a prograde orbit with $\lambda=-15_{-44}^{+50}$ deg. Although the reported error bars are relatively large, we show in Appendix \ref{app:corner} that the posteriors for the sky-projected obliquity are restricted between about -80 and 80 deg. Thus, the data strongly prefer a prograde orbit.

\begin{deluxetable*}{llcr}
\tablecaption{Summary of priors and posteriors of the joint transit photometry and out-of-transit RVs fit. $\mathcal{U}(a,b)$ denotes a uniform prior with a start value $a$ and end value $b$. $\mathcal{G}(\mu,\sigma)$ denotes a Gaussian prior with mean $\mu$, and standard deviation $\sigma$. $\mathcal{LU}(a,b)$ denotes a log-uniform prior with a start value $a$ and end value $b$. \label{tab:fit}}
\tablewidth{70pt}
%\tabletypesize{\scriptsize}
\tablehead{Parameter & Description & Prior & Posterior}
\startdata
$\rho_\star$ & Stellar density (g/cm$^3$) & $\mathcal{G}(1.40,0.05)$ & $1.41\pm0.04$ \\
\hline
$P$ & Orbital period (days) & $\mathcal{U}(104.61493,104.61653)^a$ & $104.61611\pm0.00006$ \\
$t_0$ & Transit midpoint (BJD - 2,458,000) & $\mathcal{U}(413.8472,413.8612)^a$  & $413.8515\pm0.0008$ \\
$b$ & Impact parameter & $\mathcal{U}(0,1)$ & $0.12\pm0.08$ \\
$i$ & Orbital inclination (deg) & ... & $89.93\pm0.05$ \\ 
$R_p/R_\star$ & Radius ratio & $\mathcal{U}(0,1)$ & $0.0723\pm0.0004$ \\
$a/R_\star$ & Scaled semimajor axis & ... & $93.4_{-0.9}^{+0.8}$ \\
$K$ & RV semiamplitude (m/s) & $\mathcal{U}(-100,100)$ & $<3$ ($2\sigma$)\\
$\sqrt{e}\sin{\omega}$ & Eccentricity parameter sine component & $\mathcal{U}(-1,1)$ & $-0.13_{-0.06}^{+0.05}$ \\
$\sqrt{e}\cos{\omega}$ & Eccentricity parameter cosine component & $\mathcal{U}(-1,1)$ & $-0.09_{-0.45}^{+0.40}$\\
$e$ & Eccentricity & ... & $0.09_{-0.06}^{+0.26}$ \\
$\omega$ & Argument of periastron (deg) & ... & $-123_{-41}^{+105}$ \\
$M_p$ & Planet mass ($M_\oplus$)& ... & $<20$ ($2\sigma$) \\
$R_p$ & Planet radius ($R_\oplus$)& ... & $8.2\pm0.1$ \\
$\rho_p$ & Planet mean density (g/cm$^3$)& ... & $<0.2$ ($2\sigma$)\\
$a$ & Semimajor axis (au)& ... & $0.456\pm0.005$ \\
$\dot{\gamma}$ & RV linear trend (m/s/day) & $\mathcal{U}(-1,1)$ & $-0.12\pm0.02$ \\
$\ddot{\gamma}$ & RV quadratic trend (m/s/day/yr) & $\mathcal{U}(-1,1)$ & $0.0180\pm0.0035$ \\
\hline
$\gamma_{\rm HARPS1}$ & HARPS1 RV offset (m/s)& $\mathcal{U}(25000,2700)$ & $26068\pm4$ \\
$\sigma_{\rm HARPS1}$ & HARPS1 RV jitter (m/s)& $\mathcal{LU}(10^{-3},100)$ & $10.5_{-2.2}^{+3.0}$\\
$\gamma_{\rm HARPS2}$ & HARPS2 RV offset (m/s)& $\mathcal{U}(25000,27000)$ & $26095_{-10}^{+9}$\\
$\sigma_{\rm HARPS2}$ & HARPS2 RV jitter (m/s)& $\mathcal{LU}(10^{-3},100)$ & $8.8_{-1.8}^{+2.2}$\\
$\gamma_{\rm FEROS}$ & FEROS RV offset (m/s)& $\mathcal{U}(25000,27000)$ & $26065_{-8}^{+7}$ \\
$\sigma_{\rm FEROS}$ & FEROS RV jitter (m/s)& $\mathcal{LU}(10^{-3},100)$ & $18.7_{-5.7}^{+7.7}$ \\
$\gamma_{\rm CORALIE}$ & CORALIE RV offset (m/s)& $\mathcal{U}(25000,27000)$ & $26074\pm9$ \\
$\sigma_{\rm CORALIE}$ & CORALIE RV jitter (m/s)& $\mathcal{LU}(10^{-3},100)$ & $0.1_{-0.1}^{+2.3}$\\
\hline
$q_1^{\rm TESS}$ & TESS linear limb darkening parameter & $\mathcal{U}(0,1)$ & $0.43_{-0.10}^{+0.11}$\\
$q_2^{\rm TESS}$ & TESS quadratic limb darkening parameter & $\mathcal{U}(0,1)$ & $0.10_{-0.06}^{+0.08}$\\
$\sigma_{\rm TESS}$ & TESS photometric jitter (ppm) & $\mathcal{LU}(1,5\times10^7)$ & $11_{-9}^{+54}$\\
$q_1^{R}$ & $R$ linear limb darkening parameter & $\mathcal{U}(0,1)$ &$0.53_{-0.06}^{+0.07}$ \\
$q_2^{R}$ & $R$ quadratic limb darkening parameter & $\mathcal{U}(0,1)$ & $0.92_{-0.08}^{+0.06}$\\
$\sigma_{\rm ASTEP}^{2023-R}$ & ASTEP 2023 $R$ photometric jitter (ppm) & $\mathcal{LU}(1,5\times10^7)$ & $3070_{-78}^{+75}$ \\
$\sigma_{\rm ASTEP}^{2025-R}$ & ASTEP 2025 $R$ photometric jitter (ppm) & $\mathcal{LU}(1,5\times10^7)$ & $8976_{-91}^{+90}$ \\
$q_1^{B}$ & $B$ linear limb darkening parameter & $\mathcal{U}(0,1)$ & $0.23_{-0.15}^{+0.22}$\\
$q_2^{B}$ & $B$ quadratic limb darkening parameter & $\mathcal{U}(0,1)$ & $0.15_{-0.11}^{+0.23}$\\
$\sigma_{\rm ASTEP}^{2023-B}$ & ASTEP 2023 $B$ photometric jitter (ppm) & $\mathcal{LU}(1,5\times10^7)$ & $2257_{-99}^{+103}$ \\
\enddata
\tablecomments{$^a$ The priors on the period and mid-transit time are uniform between the value $\pm$ 10$\sigma$ based on values derived in Section \ref{sec:phot}.}
\end{deluxetable*}

\section{Discussion}

\subsection{TOI-4507 b in context}

\begin{figure*}[t!]
    \centering
    \includegraphics[width=\textwidth]{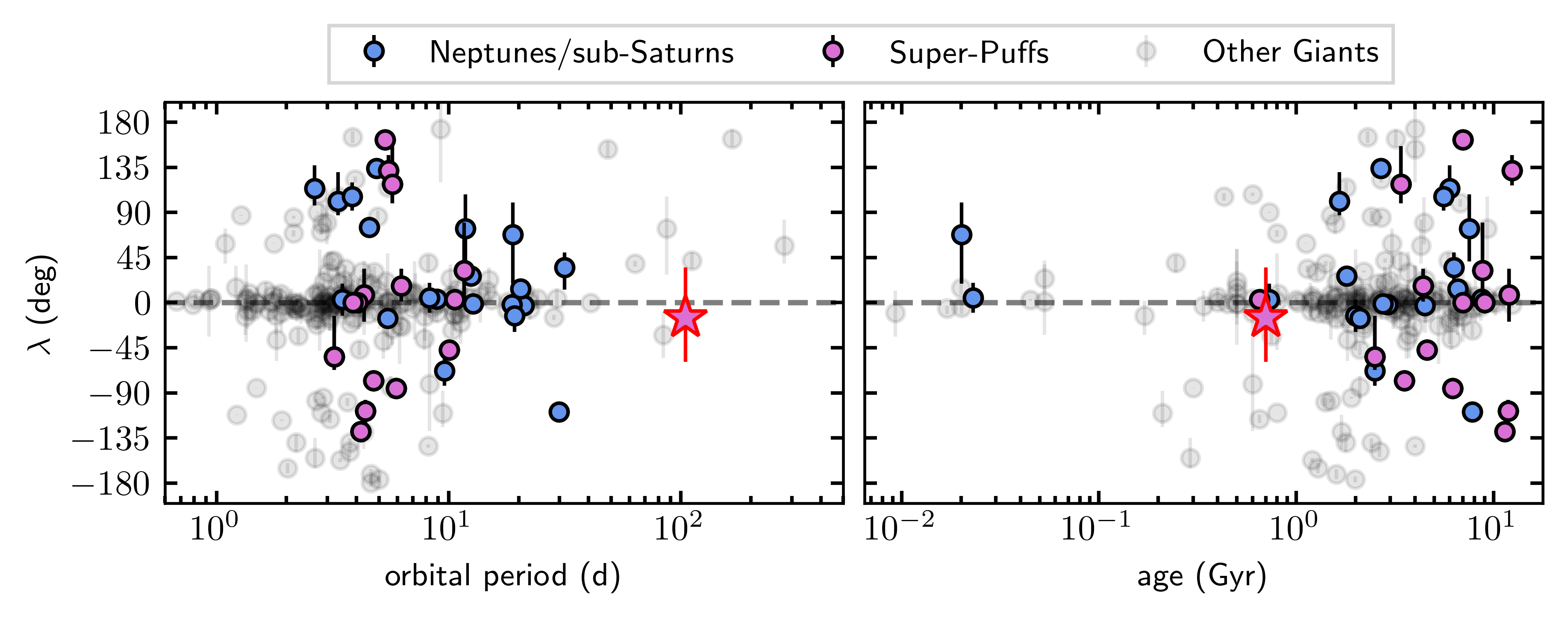}
    \caption{Sky-projected obliquity $\lambda$ as a function of the orbital period (left) and age of the system (right). The population of Neptunes/sub-Saturns ($10<M_p/M_\oplus<90$) is shown in blue. Super-puff planets (Neptunes/sub-Saturns with $\rho_p<0.3$ g/cm$^3$) are shown in pink, with TOI-4507 highlighted as a star with a red edge. Other giant planet systems are shown as faint black points. Data from TEPCat with ages from the NASA Exoplanet Archive.}
    \label{fig:population}
\end{figure*}

TOI-4507 b appears to be an interesting planet in several ways. Figure~\ref{fig:population} shows the results for the projected obliquity $\lambda$ along with those of other systems. The obliquity data were drawn from TEPCat\footnote{\url{https://www.astro.keele.ac.uk/jkt/tepcat/}} \citep{Southworth2011} and the age data were drawn from the NASA Exoplanet Archive \citep{Akeson2013,Christiansen2025}, both as of May 2025. Evidently, TOI-4507 b has one of the longest orbital periods for which $\lambda$ has been measured, having the fourth-longest orbital period after HIP 41378 d \citep{Grouffal2022}, TIC 241249530 b \citep{Gupta2024}, and HD 80606 b \citep[e.g.,][]{Hebrard2010}. Here, we define Neptunes/sub-Saturns as the class of planets with $10<M_p/M_\oplus<90$, with the subclass of super-puffs having $\rho_p<0.3$ g/cm$^3$.

TOI-4507 b  joins a growing population of Neptune/sub-Saturn systems with measured stellar obliquities \citep[e.g.,][]{Bourrier2023,Espinoza-Retamal2024,Knudstrup2024,Handley2025}. This population includes some other super-puffs on prograde orbits, such as K2-261 b \citep{Knudstrup2024}, TOI-5398 b \citep{Mantovan2024,Radzom2024}, WASP-21 b \citep{Chen2020}, WASP-39 b \citep{Mancini2018}, WASP-69 b \citep{Casasayas-Barris2017}, WASP-117 b \citep{Lendl2014}, and WASP-193 b \citep{Yee2025}. Within this collection of prograde super-puffs, TOI-4507 b has the longest period.

\begin{deluxetable*}{llcr}
\tablecaption{Summary of priors and posteriors of the RM effect fit. $\mathcal{U}(a,b)$ denotes a uniform prior with a start value $a$ and end value $b$. $\mathcal{G}(\mu,\sigma)$ denotes a normal prior with mean $\mu$, and standard deviation $\sigma$. $\mathcal{LU}(a,b)$ denotes a log-uniform prior with a start value $a$ and end value $b$. \label{tab:RM}}
\tablewidth{70pt}
\tablehead{Parameter & Description & Prior & Posterior}
\startdata
$\lambda$ & Sky-projected stellar obliquity (deg) & $\mathcal{U}(-180,180)$ & $-15_{-44}^{+50}$ \\
$v\sin{i_\star}$ & Projected rotational velocity (km/s) & $\mathcal{U}(0,15)$ & $7.3_{-1.8}^{+2.8}$ \\
$\rho_\star$ & Stellar density (g/cm$^{3}$) & $\mathcal{G}(1.41,0.04)$ & $1.41\pm0.04$ \\
\hline
$P$ & Orbital period (days) & $\mathcal{G}(104.61611,0.00006)$ & $104.61611\pm0.00006$ \\ 
$t_0$ & Transit midpoint (BJD) & $\mathcal{G}(2458413.8515,0.0008)$ & $2458413.8514\pm0.0008$ \\
$b$ & Impact parameter & $\mathcal{G}(0.12,0.08)$ & $0.11_{-0.06}^{+0.07}$ \\
$R_p/R_\star$ & Radius ratio & $\mathcal{G}(0.0723,0.0004)$ & $0.0723\pm0.0004$ \\
$\sqrt{e}\sin{\omega}$ & Eccentricity parameter sine component & $\mathcal{G}(-0.13,0.06)$ & $-0.15\pm0.06$ \\
$\sqrt{e}\cos{\omega}$ & Eccentricity parameter cosine component & $\mathcal{G}(-0.09,0.45)$ & $-0.08_{-0.37}^{+0.42}$ \\
$e$ & Eccentricity & ... & $0.11_{-0.08}^{+0.20}$ \\
$\omega$ & Argument of periastrom & ... & $-117_{-46}^{+96}$ \\
\hline
$q_1^{\rm HARPS}$ & HARPS linear limb darkening parameter & $\mathcal{U}(0,1)$ & $0.59_{-0.31}^{+0.27}$ \\
$q_2^{\rm HARPS}$ & HARPS quadratic limb darkening parameter & $\mathcal{U}(0,1)$ & $0.65_{-0.36}^{+0.25}$ \\
$\beta$ & Intrinsic stellar line width (km/s) & $\mathcal{G}(6.0,2.0)$ & $5.9_{-2.1}^{+2.0}$ \\
$\gamma_{\rm HARPS}$ & HARPS RV offset (m/s)& $\mathcal{U}(25000,27000)$ &  $26068_{-10}^{+9}$ \\
$\dot{\gamma}_{\rm HARPS}$ & HARPS RV slope (m/s/day)& $\mathcal{U}(-200,200)$ & $22_{-63}^{+61}$ \\
$\sigma_{\rm HARPS}$ & HARPS RV jitter (m/s)& $\mathcal{LU}(10^{-3},100)$ & $1.8_{-1.8}^{+3.7}$ \\
\enddata
%\tablecomments{}
\end{deluxetable*}

Additionally, with an age of $\sim700$ Myr, TOI-4507 b is the youngest super-puff with an obliquity measurement. Further, it is one of the youngest Neptunes/sub-Saturns in general with an obliquity constraint, only behind AU Mic c \citep{Yu2025}, V1298 Tau c \citep{Feinstein2021}, and TOI-5398 b \citep{Mantovan2024,Radzom2024}.

TOI-4507 b seems well-suited for atmospheric characterization via transit spectroscopy. Assuming zero albedo, the expected equilibrium temperature of the planet is $\sim460$ K, lower than most other systems that have been subject to transit spectroscopy. Because of the planet's low density, the transmission spectroscopy metric \citep[TSM;][]{Kempton2018} is $\gtrsim160$. For reference, targets with TSM > 90 are typically considered high-quality targets for atmospheric characterization. Observations with JWST might shed light on the atmospheric composition of TOI-4507 b and the origins of its unusually low density.

\subsection{Search for companions}

Although there are no signs of additional transiting planets in the TESS light curve, there is evidence for a long-term RV trend. To discern whether this trend originates from an outer companion or is just a product of stellar activity, we calculated a series of stellar activity indicators using \texttt{serval} \citep{serval} and the HARPS data, including the H$\alpha$ index, the Na D I and II indices, the chromatic index, and the differential line width. The H$\alpha$ index is the only one showing a (marginal) correlation with the derived RVs, having a Pearson correlation coefficient $r = 0.35 \pm 0.13$ and a $p{\rm -value} = 0.053^{+0.19}_{-0.05}$. Therefore, the hypothesis of stellar activity is disfavored. 

If the RV trend is produced by a planetary companion, it should have an orbital period $P>5.6$ yr, which corresponds to the baseline of the current RV observations. The parameters of this possible companion could be constrained with additional RV and/or {\it Gaia} DR4 and DR5 astrometric measurements. Although TOI-4507 is relatively massive, it is nearby ($d\approx177$ pc) and bright ($G\approx10.6$), so it is a good candidate for {\it Gaia} astrometric detections \citep[e.g.,][]{Perryman2014,Lammers2025}. When removing the signal of TOI-4507 b and the RV trend from the data, we do not see clear additional signals in the RV data, which rules out the presence of planets producing similar or larger RV signals than TOI-4507 b.

In addition to planetary companions, we searched for potential stellar companions to TOI-4507. High-resolution imaging with SOAR rules out nearby sources within $3^{\prime\prime}$ (Figure~\ref{fig:soar}), corresponding to a projected separation of $\sim530$~au. This is further supported by the {\it Gaia} DR3 Renormalized Unit Weight Error of $\sim0.8$, which suggests that the astrometry is not significantly perturbed by an unresolved companion. On wider scales, TOI-4507 is not included in the wide binary catalog of \citet{El-Badry2021}, which is based on proper motions and parallaxes from the {\it Gaia} Early DR3 \citep{Gaia_EDR3}, indicating the absence of co-moving stellar companions within 1 pc. The available data then strongly support a single-star interpretation.

\subsection{Explaining the Inferred Low Density}

Next, we discuss some of the proposed explanations for the low-density planets.

\paragraph{Composition}
One of the suggested hypotheses to explain the low densities of some exoplanets is that they have unusually massive H/He envelopes relative to their small core masses, typically representing more than $\sim20\%$ of their total masses \citep[e.g.,][]{Lopez2014,Thorngren2016}. \citet{Lee2016} hypothesize that super-puff exoplanets might form in regions of their protoplanetary disks that are unusually cold and dust-free, and then migrate to their currently observed orbits. This picture can explain their super-puffy nature, and also the fact that many super-puffs are in or close to mean-motion resonances \citep[e.g.,][]{Leleu2024} if they experience slow convergent migration. Future observations of TOI-4507 with JWST can probe the atmospheric composition of the planet, which might confirm or rule out this hypothesis.

\paragraph{Tidal heating}
An alternative hypothesis is that super-puff planets are inflated due to tidal heating produced by distortions in the shape of the planet. These distortions are more pronounced the closer the planet orbits the star, as tidal forces become stronger \citep[e.g.,][]{Hut1981}. \citet{Millholland2019} and \cite{Millholland2020} showed that for planets in the Neptune regime, the radius can increase by up to a factor of $\sim2$, possibly explaining the super-puffs. Indeed, most of the discovered super-puffs have short-period orbits and experience strong tidal forces, making this a valid explanation for them. However, this is not the case for TOI-4507 b, which has an orbital period of $105$ days and does not experience significant tidal forces, making this scenario unlikely to explain its unusually low density.

\paragraph{Circumplanetary rings}
One plausible explanation for the inferred super-puffy nature of some exoplanets with relatively long orbital periods is the idea of opaque rings around the planets \citep[e.g.,][]{Piro2020,Akinsanmi2020,Saillenfest2023}. Recently, \citet{Lu2025} showed that an optically thick ring system around HIP-41378 f could remain tilted due to the planet's resonantly excited obliquity, thus explaining its unusually deep transit.

This scenario requires that the optically thick ring be both sustainable against Poynting-Robertson (PR) drag and maintained in a high-obliquity state. Regarding the first condition, \citet{OhnoFortney2022} derived a critical equilibrium temperature above which a viscously evolving ring cannot persist:
\begin{equation}
T_{\rm crit}\sim 300~{\rm K}\left(\frac{{\rm Age}}{0.7\;{\rm Gyr}}\right)^{-3/8}.    
\end{equation}
Since the planet’s equilibrium temperature of $T_{\rm eq}\approx 460$~K is well above this threshold, the ring is unlikely to survive over the system’s lifetime.

The second condition concerns the ring’s obliquity. The radial extent over which a ring remains aligned with the planetary equator is set by the Laplace radius ($R_L$). For a planet with a rotationally induced quadrupolar moment $J_2$, this is
\begin{equation}
R_L\simeq\frac{2R_p}{3}\left(\frac{k_{2,p}}{0.4}\right)^{1/5}
\left(\frac{P}{P_{\rm rot,p}}\right)^{2/5},
\end{equation}
where $P_{\rm rot,p}$ is the planetary spin period and $k_{2,p}$ is the Love number. For a planetary obliquity $\psi_{\rm p}$, the observed transit depth requires a projected ring area satisfying $R_L\sin \psi_{\rm p} \gtrsim 8\,R_\oplus$. In terms of the spin period, this condition becomes
\begin{equation}
P_{\rm rot}\lesssim 1.1~{\rm days}\,
\left(\frac{R_{p}}{4R_{\oplus}}\frac{1}{\sin \psi_{\rm p}}\right)^{5/2}.
\end{equation}
Although sub-day rotation periods may be reasonable, or even expected, for a Neptune-like planet, the required values for rocky planets may approach break-up rotation period (e.g., $P_{\rm rot}\lesssim 1$ h for $R_p\lesssim 1.1\,R_\oplus$). Assuming instead a Neptune-like planet, the rotation must remain rapid despite tidal damping. The corresponding spin-down timescale is (e.g., \citealt{correia2010}):

\begin{eqnarray}\label{eq:tau_spin}
\tau_{\rm spin}&\equiv&\frac{P_{\rm rot,p}}{|\dot{P}_{\rm rot,p}|} \approx
0.7\times 10^9~{\rm yr}\,
\left(\frac{Q_p/k_{2,p}}{10^5}\right)
\left(\frac{M_p}{20M_\oplus}\right)\nonumber\\
&\times&\left(\frac{R_p}{4R_\oplus}\right)^{-3}
\;\left(\frac{2}{1+\cos^2\psi_{\rm p}}\right),
\end{eqnarray}
which must exceed the system age ($\sim 0.7$ Gyr). This condition implies a modified tidal quality factor $Q_p/k_{2,p}\gtrsim 10^5$, somewhat higher than the values inferred for Neptune and Uranus of $\sim 10^4$ (e.g., \citealt{GS1996}), but within the range of warm Neptunes based on eccentricity damping constraints (e.g., \citealt{correia2020}).

In summary, the primary challenge for the ring hypothesis is that long-lived opaque rings are unlikely to survive at the planet’s equilibrium temperature. If such rings were present, sustaining them in a high-obliquity configuration would require the planet to have a modified tidal quality factor of $Q_p/k_{2,p}\gtrsim 10^5$, a value that may be somewhat elevated but still plausible for a Neptune-like planet.

\section{Conclusions}

We have reported the discovery and characterization of TOI-4507 b, a transiting cold Neptune on a prograde orbit. With a radius of $8.2\pm0.1\,R_\oplus$, a mass $<20\,M_\oplus$, and a bulk density $<0.2$ g/cm$^3$, TOI-4507 b is among the longest-period ($P\sim105$ d) super-puffs discovered to date. Further, TOI-4507 is one of the youngest Neptune/sub-Saturn systems with an obliquity constraint, and one of the longest-period systems in general for which the obliquity has been measured.

The properties of TOI-4507 b raise important implications. Its unusually low density highlights the need to consider both compositional and structural explanations for super-puff planets, including massive H/He envelopes, inflation mechanisms, or obscuration by circumplanetary material. However, our analysis suggests that long-lived opaque rings are unlikely to survive at the planet’s equilibrium temperature.

Given its brightness, young age, and extremely low density, TOI-4507 b is a prime target for follow-up studies. Transmission spectroscopy with JWST can test the super-puff hypothesis by probing its atmospheric composition, while continued RV and astrometric monitoring can confirm the presence of an outer companion and further constrain the dynamical history of the system.

TOI-4507 thus joins a small but growing population of Neptune/sub-Saturn and super-puff systems with obliquity constraints, providing a valuable laboratory for understanding the formation, evolution, and diversity of planetary systems.

\begin{acknowledgments}

We would like to thank Yubo Su for suggesting the rotational constraint on the sizes of the rings. We also thank Yugo Kawai for kindly pointing out that the source of the variability observed in the TESS light curve of TOI-4507 is a blended variable star in the background, and it is not related to the rotational period of TOI-4507. We would also like to thank the anonymous referee for their thoughtful review and suggestions that improved the quality of this work. We also thank Songhu Wang, Xian-Yu Wang, and the Red Worlds Lab group at the University of Amsterdam for their useful discussions and suggestions that strengthened this manuscript.

J.I.E.-R. gratefully acknowledges support from the John and A-Lan Reynolds Faculty Research Fund, from the ANID BASAL project FB210003, and from the ANID Doctorado Nacional grant 2021-21212378.
R.B. acknowledges support from FONDECYT Project 1241963 and from ANID -- Millennium  Science  Initiative -- ICN12\_009.
T.T. acknowledges support by the BNSF program ``VIHREN-2021" project No. KP-06-DV/5.
Support for M.C. is provided by ANID’s Millennium Science Initiative through grants ICN12\_12009 and AIM23-001, awarded to the Millennium Institute of Astrophysics (MAS); by ANID/FONDECYT Regular grant 1231637; and by ANID/Basal (CATA) grant FB21003. 
C.C. acknowledges support from ANID through FONDECYT post-doctoral grant number 3230283.
A.V.F. acknowledges the support of the Institute of Physics through the Bell Burnell Graduate Scholarship Fund.

This paper was based on observations collected at the European Southern Observatory under ESO programmes 0104.C-0413(A), 106.21ER.001, 112.25W1.001, 114.27CS.001, and 115.286G.001. This paper also made use of data collected by the TESS mission and are publicly available from the Mikulski Archive for Space Telescopes (MAST) operated by the Space Telescope Science Institute (STScI). Funding for the TESS mission is provided by NASA's Science Mission Directorate. We acknowledge the use of public TESS data from pipelines at the TESS Science Office and at the TESS Science Processing Operations Center. Resources supporting this work were provided by the NASA High-End Computing (HEC) Program through the NASA Advanced Supercomputing (NAS) Division at Ames Research Center for the production of the SPOC data products. This work makes use of observations from the ASTEP telescope. ASTEP benefited from the support of the French and Italian polar agencies IPEV and PNRA in the framework of the Concordia station program and from OCA, INSU, Idex UCAJEDI (ANR- 15-IDEX-01) and ESA through the Science Faculty of the European Space Research and Technology Centre (ESTEC). This research also received funding from the European Research Council (ERC) under the European Union's Horizon 2020 research and innovation program (grant agreement No. 803193/BEBOP), from the Science and Technology Facilities Council (STFC; grant No. ST/S00193X/1, ST/W002582/1 and ST/Y001710/1), and from the ERC/UKRI Frontier Research Guarantee programme (CandY/ EP/Z000327/1). This work has been carried out within the framework of the National Centre of Competence in Research PlanetS supported by the Swiss National Science Foundation under grants 51NF40\_182901 and 51NF40\_205606. This paper is also based on observations obtained at the SOAR telescope, which is a joint project of the Minist\'{e}rio da Ci\^{e}ncia, Tecnologia e Inova\c{c}\~{o}es (MCTI/LNA) do Brasil, the US National Science Foundation’s NOIRLab, the University of North Carolina at Chapel Hill (UNC), and Michigan State University (MSU). %This work has been carried out within the framework of the National Centre of Competence in Research PlanetS supported by the Swiss National Science Foundation.
This research has made use of the Exoplanet Follow-up Observation Program (ExoFOP) and NASA Exoplanet Archive websites, which are operated by the California Institute of Technology under contract with NASA under the Exoplanet Exploration Program. 

\facilities{TESS, ASTEP, SOAR, ESO:3.6m (HARPS), Max Planck:2.2m (FEROS), Euler1.2m (CORALIE), Exoplanet Archive, ExoFOP.}

\software{
\texttt{astropy} \citep{astropy,astropy2,astropy3},
\texttt{batman} \citep{batman},
\texttt{ceres} \citep{ceres},
\texttt{celerite} \citep{celerite},
\texttt{dynesty} \citep{dynesty},
\texttt{ironman} \citep{Espinoza-Retamal2023b,Espinoza-Retamal2024},
\texttt{juliet} \citep{juliet},
\texttt{lightkurve} \citep{lightkurve},
\texttt{radvel} \citep{radvel},
\texttt{rmfit} \citep{Stefansson22},
\texttt{serval} \citep{serval},
\texttt{zaspe} \citep{zaspe}.
}

\end{acknowledgments}

\bibliography{sample7}{}
\bibliographystyle{aasjournalv7}

\appendix
\restartappendixnumbering

\section{Radial Velocity Measurements}\label{app:rvs}

Table \ref{tab:rv} shows the radial velocities of TOI-4507 taken with different instruments.

\begin{deluxetable}{cccc}[h!]
\digitalasset
\tablecaption{Radial velocity measurements from HARPS, FEROS, and CORALIE.}\label{tab:rv}
\tablehead{
   \colhead{BJD} & \colhead{RV (m/s)} & \colhead{$\sigma_{\rm RV}$ (m/s)} & \colhead{Instrument}
}
\startdata
2458717.932107 & 26094.5 & 11.0 & FEROS\\
2458718.898281 & 26072.7 & 9.0 & FEROS\\
2458722.879818 & 26053.6 & 9.3 & FEROS\\
\nodata & \nodata & \nodata & \nodata \\
2460667.658069 & 26062.9 & 2.3 & HARPS\\
2460692.667466 & 26060.0 & 3.8 & HARPS\\
2460744.564697 & 26059.6 & 3.3 & HARPS\\
\enddata
\tablecomments{Table \ref{tab:rv} is published in its entirety in the machine-readable format. A portion is shown here for guidance regarding its form and content.}
\end{deluxetable}

\clearpage

\section{Source of the Photometric Variability}\label{app:contamination}

\begin{figure*}[h!]
    \centering
    \includegraphics[width=0.9\textwidth]{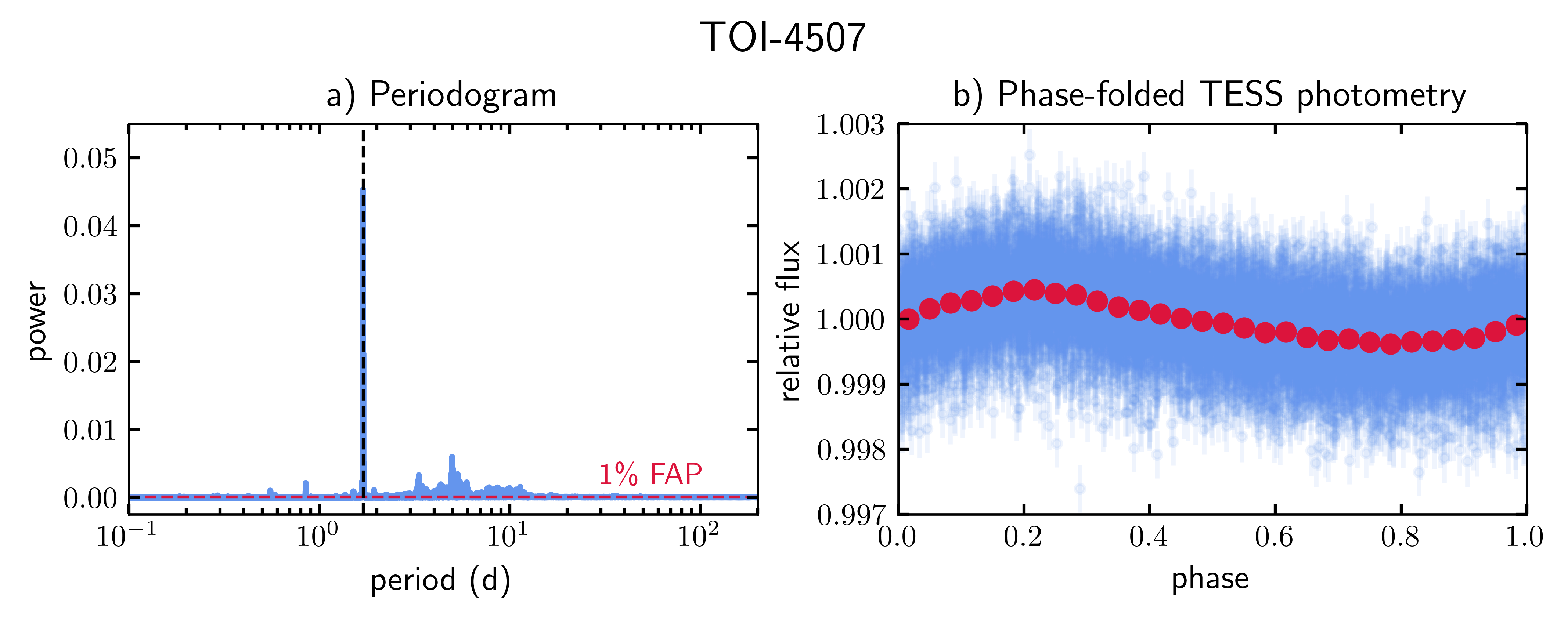}
    \includegraphics[width=0.9\textwidth]{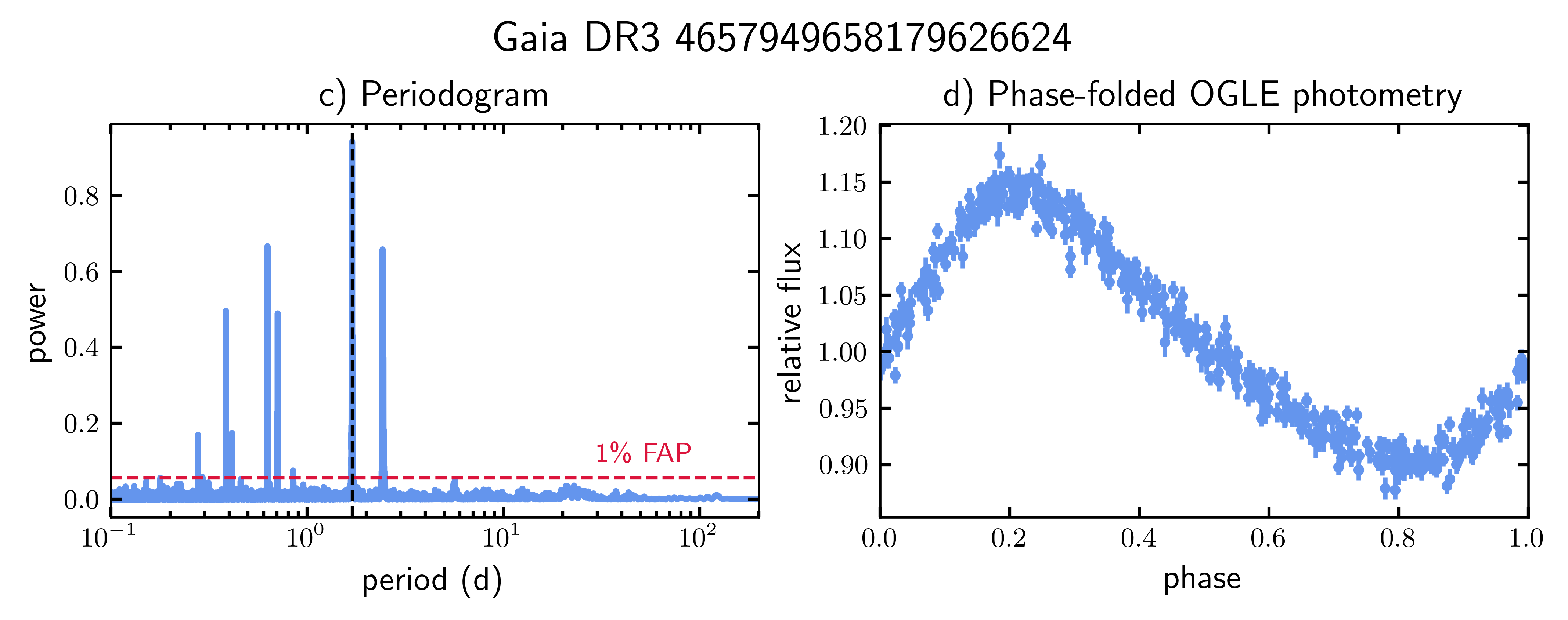}
    \caption{Observed modulation in the light curves of TOI-4507 (upper panels) and the nearby source Gaia DR3 4657949658179626624 (bottom panels). a) Lomb-Scargle periodogram of the 2-min cadence TESS light curve of TOI-4507. A false alarm probability (FAP) level of 1\% is marked with a red dashed line. The black dashed line marks the highest-power peak, indicating a period of 1.7 days. b) TESS light curve phased to the period of 1.7 days. All the work done here made use of the 2-minute cadence light curve, but for illustrative purposes only, the 30-minute cadence TESS light curve is shown here in blue, with binned data in red. c) Lomb-Scargle periodogram of the OGLE light curve of Gaia DR3 4657949658179626624, also showing a clear period of 1.7 days. d) OGLE light curve of Gaia DR3 4657949658179626624 phased to the period of 1.7 days.}
    \label{fig:Prot}
\end{figure*}

As discussed in Section \ref{sec:phot}, the TESS light curve of TOI-4507 shows a clear modulation with periodicity of $\sim1.7$ days (see the upper panels in Figure \ref{fig:Prot}). To investigate whether this variability is intrinsic to TOI-4507 (e.g., due to stellar rotation) or the result of contamination from blended sources in the TESS pixels, we checked archival Optical Gravitational Lensing Experiment \citep[OGLE;][]{Udalski1992} data of nearby sources. We identified a variable star, Gaia DR3 4657949658179626624, located at $\sim32^{\prime\prime}$ of TOI-4507, that shows variability with the same periodicity but with a larger amplitude (see the bottom panels in Figure \ref{fig:Prot}). The variability amplitude of this neighboring source would translate to a variation consistent with the modulation observed in the blended TESS light curve. We then conclude that the variability observed in TESS is not related to TOI-4507 or its rotational period. Therefore, we are unable to constrain the stellar inclination and true 3D obliquity in the analysis.

\clearpage

\section{Corner Plot}\label{app:corner}

\begin{figure*}[h!]
    \centering
    \includegraphics[width=\linewidth]{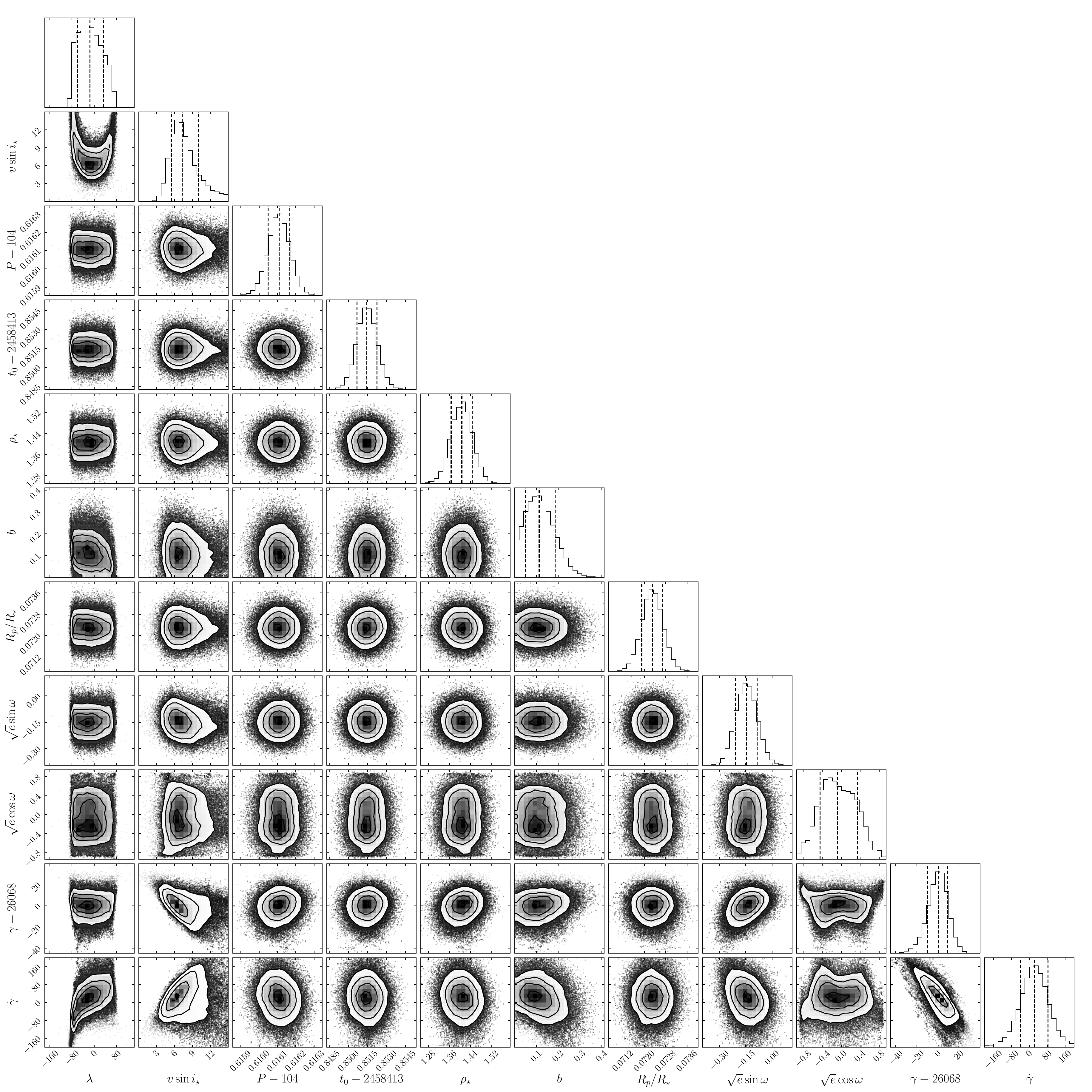}
    \caption{Joint posterior distributions and histograms of the posteriors for some of the parameters in our RM effect model.}
    \label{fig:corner}
\end{figure*}

Figure \ref{fig:corner} shows the joint posterior distributions and histograms for some of the parameters of the RM fit described in Section \ref{sec:RM}.

\end{document}